\documentclass[aps,showpacs,a4paper,floatfix,twocolumn,prc,amsmath,amssymb]{revtex4-1}
\usepackage{amsmath}
\usepackage{graphicx}
\usepackage{ulem}
\usepackage{morefloats}
\usepackage{rotating}
\usepackage{relsize}
\usepackage{color}
\usepackage{floatflt,epsfig}
\usepackage{float}
\usepackage{ulem}
\usepackage{natbib}
\usepackage{hyperref}
\usepackage{filecontents}

\begin{document}

\title{Anisotropic Compact Stars in Rastall-Rainbow Gravity}

\author{Cl\'esio E. Mota $^1$}
\email{clesio200915@hotmail.com}
\author{Luis C. N. Santos$^2$}
\email{luis.santos@ufsc.br}
\author{Franciele M. da Silva$^1$}
\email{franmdasilva@gmail.com}
\author{Cesar V. Flores$^1$}
\email{cesarovfsky@gmail.com}
\author{Tiago J. N. da Silva$^1$}
\email{t.j.nunes@ufsc.br}
\author{D\'ebora P. Menezes $^1$}
\email{debora.p.m@ufsc.br}

\affiliation{$^1$Departamento de F\'isica, CFM - Universidade Federal de Santa Catarina; C.P. 476, CEP 88.040-900, Florian\'opolis, SC, Brasil}
\affiliation{$^2$Departamento de F\'isica, CCEN - Universidade Federal da \\ Para\'iba; C.P. 5008, CEP  58.051-970, Jo\~ao Pessoa, PB, Brasil.}

\begin{abstract}

In this work, we have investigated anisotropic neutron stars in the framework of Rastall-Rainbow gravity. All our calculations were computed using the IU-FSU realistic equation of state (EoS), in which was considered two cases: standard nucleonic composition and the one with the eight lightest baryons. From the neutron star masses and radii obtained we conclude that anisotropic pressure has significant consequences on the structure of stellar objects. In particular, when anisotropy is considered within the general relativity framework, it significantly modifies the maximum stellar mass. On the other hand, when Rastall-Rainbow gravity and anisotropy are simultaneously considered, they provide the best results for mass and radius values, including important astrophysical objects such as the LMXB NGC 6397 and the extremely massive pulsar millisecond MSP J0740 + 6620. Although the expected inclusion of hyperons in the nuclear model reproduces stellar masses smaller than those produced by standard nucleonic matter, we shown that the hyperon puzzle problem can be solved by including anisotropic effects on compact stars in the context of the Rastall-Rainbow gravity.

\noindent Keywords : general relativity, Rastall-Rainbow gravity, neutron stars. 
\end{abstract}
\maketitle
\section{Introduction}

General Relativity (GR) is widely accepted as the theory of gravity. Since its proposition, GR has changed our understanding of the nature of space and time and has passed a substantial number of experimental tests. Among these tests, lies the great precision with which it predicts the orbital precession of the mercury perihelion, the recent detection of gravitational waves generated by binary neutron star (NS) systems \cite{PhysRevLett.119.161101} and the first observation of a black hole shadow obtained by the Event Horizon Telescope \cite{2019ApJ...875L...1E}. 

However, although predictive in a wide range of situations, GR has some limitations, and there are still some questions that it cannot explain satisfactorily, either in the context of cosmology or in the astrophysical context, as for instance, the dark matter problem, the dark energy problem, and the early inflation \cite{fder1,Resco2016}. These limitations have lead to the proposal of the so-called modified theories of gravity, which extend GR through the introduction of additional terms in the integrand of the Einstein-Hilbert action. Examples of those include the $f(R)$ gravity and extensions (see \cite{Harko2018} for an extensive review). 

In addition to the modified theories of gravity discussed above, we also mention the Rastall  \cite{Rastall} and Rainbow theories \cite{magueijo}, built by changing some fundamental conceptions of gravity and principles of GR. These two theories have been combined in a recent work and shown to provide a good description of both massive and low radii NS \cite{mota2019}.

The limits of the validity of GR are being probed by important experimental tests in regions of extreme gravity. Comprehensive efforts have been concentrated, for instance, on the detection of gravitational waves emitted by mergers of compact objects such as neutron stars and black holes. In fact, the GR prediction that gravitational waves travel at the speed of light has been recently verified by the multi-messenger observation of the GW170817 event \cite{PhysRevLett.119.161101}, which has also excluded some of the proposed theories of modified gravity.

In addition, neutron stars (NS) are considered important astrophysical laboratories that allow for different theoretical models of nuclear matter under extreme physical conditions and alternative gravity theories \cite{Eksi2014,Folomeev2018} to be tested. Gravity modification is expected to affect a number of important physical characteristics of NS, such as mass and radius, for which observations are becoming more and more accurate. The launching of the NICER telescope \cite{NASANICE} in 2017 promised precise values for masses and radii in a near future.

From the theoretical/phenomenological point of view, one needs to combine modern equations of state which model neutron nuclear matter with an appropriate description of gravity. When studying the physical characteristics of NS it is usual to consider that the matter content within the star is described by an isotropic perfect fluid equation of state, which possesses equal radial and tangential pressure components. On the other hand, some astronomers believe that magnetic fields (B) directly influence the formation of the structure of NS as much as the gravitational attraction. Usually these objects have a magnetic field of the order of $10^{12}$ G and are generally detected in the universe as pulsars, but some present even stronger magnetic fields \cite{Lopes2015} and are commonly called magnetars. In these cases, anisotropic effects are expected due to tensions in stellar matter \cite{Folomeev2018,Chaichian1999,PerezMartinez2007,Ferrer2010}. On the other hand, regardless of the physical nature of the appearance of anisotropy in stellar fluid, its effects lead to significant changes in the star structure \cite{Bowers1974,Heintzmann1975,Hillebrandt1976,Bayin1982,Bondi1992,Herrera1997,Mak2001,Horvat2010,Herrera2013}. The necessity of dealing or not with anisotropic pressure and how to compute its effects are still controversial issues, but there are a number of different prescriptions in the literature and some recent discussions can be found in \cite{Menezes2015,Lopes2015,Debi2015,Debi2019,Lopes2019}. Within this context and motivated by recent observations of NS with masses above 2.0 M$_{\odot}$ \cite{Cromartie2019}, we next study anisotropic compact stars in the framework of the Rastall-Rainbow model of  gravity (RR), as proposed in \cite{mota2019}. 

This work is divided as follows: in section II we review the Rastall-Rainbow model and the resulting TOV-like equations for anisotropic compact stars are obtained; we present our main results and discussions in section III and draw the final conclusions in section IV.

\section{Rastall-Rainbow gravity and stellar structure equations}

\subsection{Rastall-Rainbow theory}
\label{TOV_RR}

The Rastall-Rainbow gravity model \cite{mota2019} comes from the combination of two modified theories  of gravity that extend GR, namely the Rastall theory \cite{Rastall} and the Rainbow theory \cite{magueijo}.

In 2004, Jo\~ao Magueijo and Lee Smolin \cite{magueijo} proposed a generalization of nonlinear (or deformed) special relativity to curved space-time. They have shown how the formalism, which characterizes the principles of this relativity, can be generalized to incorporate curvature, leading to what may be called {\it double general relativity}. Initially, they proposed a dual realization of this nonlinear relativity in the momentum space, and thus demonstrated that for such case, the space-time invariant becomes an energy dependent metric. Rainbow gravity is caused by the modification of the usual relativistic relation $E^{2} - p^{2} = m^{2}$ in the high energy regime. This modification is induced by two arbitrary functions (Rainbow functions) $ \Xi(x)$ and $\Sigma(x)$, satisfying
\begin{equation}
    E^{2} \Xi(x)^{2}-p^{2}\Sigma(x)^{2}=m^{2},
\label{eq1}
\end{equation}
where $x = E/E_{p}$ is the relationship between the energy of the test particle $E$ 
and the Planck energy $E_{p} = \sqrt{\frac{\hslash c^{5}}{G}}$. The quantities $m$ and $p$ are the respectively the particle
mass and momentum. Consequently, $ \Xi(x)$ and $\Sigma(x)$ are chosen so that, under a low energy regime $x = E/E_{p} \rightarrow 0 $, the usual dispersion ratio is recovered so as to satisfy the relations: 
\begin{equation}
    \lim_{x\rightarrow 0} \Xi(x)=1, \quad \lim_{x \rightarrow 0} \Sigma(x)=1.
    \label{eq2}
\end{equation}
%
The space-time in this context is usually constructed using the following energy dependent metric \cite{magueijo}
\begin{equation}
    g(x)=\eta^{ab} e_{a}(x)\otimes e_{b}(x),
    \label{eq3}
\end{equation}
where $e_{a}(x)$ e and $e_{b}(x)$ are the energy dependent fields and are related to the independent fields, denoted by  $\widetilde{e}_{a}$ and $\widetilde{e}_{b}$, according to the following expressions:
\begin{equation}
    e_{0}(x)=\frac{1}{ \Xi (x)} \widetilde{e}_{0}, \quad e_{b}(x)=\frac{1}{\Sigma (x)} \widetilde{e}_{b}.
    \label{eq4}
\end{equation}
We identify $b = (1, 2, 3)$ as the spatial coordinates. Now we use Einstein's field equations modified by Rainbow gravity, under the assumption that the space-time geometry is dependent on the energy of the test particle (EPT). Thus all the quantities that make up the field equations in this gravity become dependent on this energy. In addition, Einstein's field equations are replaced by a family of other field equations, such as
\begin{equation}
G_{\mu\nu}(x) \equiv  R_{\mu\nu}(x)-\frac{1}{2}g_{\mu\nu}(x)R(x)= k(x)T_{\mu\nu}(x)  
  \label{eq10},
\end{equation}
where $k(x) = 8 \pi G(x)$. 

In 1972, Peter Rastall \cite{Rastall} maintaining the validity of the gravitational Bianchi identity of the Einstein geometric tensor $(\nabla^{\mu}G_{\mu\nu} = 0)$, proposed a generalization concerning the conservation principles of the energy momentum tensors, considering that in curved space-time we had $\nabla^{\mu}T_{\mu\nu} \neq 0$. Based on phenomenological justifications, Rastall considered that the divergence of $T_{\mu\nu}$ was proportional to the variation of the Ricci scalar $R$. He postulated that the curvature itself contributed to the total energy of the system. The modification of the conservation law of the energy-momentum tensor in curved space-time, is written as \cite{mota2019}:
\begin{equation}
 \nabla^{\mu}T_{\mu\nu} = \bar{\lambda}\nabla_{\nu}R,
 \label{eq6}
\end{equation}
where $\bar{\lambda} = \frac{1 - \lambda}{16 \pi G}$. Here, $\lambda$ is a constant called Rastall parameter, which measures the possibility that the geometry couples with the fields of matter \cite{Dipanjana}. The modification inserted in equation (\ref{eq6}) is such that for $\bar{\lambda} = 0$, or, equivalently, $\lambda = 1$, we adequately recover GR. The Ricci scalar $R$ vanishes in flat space-time, and thus the conservation law for $T_{\mu\nu}$ in this context is restored, what reinforces the obvious condition that Rastall theory imposes a non-flat space-time.

In fact, modified Einstein field equations consistent with equation (\ref{eq6}) can be written in the simplified form
\begin{equation}
R_{\mu}{}^{\nu} - \frac{\lambda}{2}\delta_{\mu}{}^{\nu}R = 8\pi G T_{\mu}{}^{\nu}, \label{eq8}
\end{equation}
which are the field equations modified by Rastall's gravity. 

The Rastall-Rainbow model unifies the effects of Rainbow gravity with the effects of Rastall gravity, under a single formalism. The field equations in this unified formalism are written as follows:
\begin{equation}
R_{\mu}{}^{\nu}(x) - \frac{\lambda}{2}\delta_{\mu}{}^{\nu}(x)R(x) = k(x) T_{\mu}{}^{\nu}(x)  
 \label{eq7},
\end{equation}
where the Rastall parameter $\lambda$ is independent of the test particle energy. In the following section, we describe the procedures for solving  equation (\ref{eq7}) in the description of stellar hydrostatic equilibrium. In fact, such an approach results in modified TOV \cite{tov1,tov2} equations due to the modifications inserted by Rastall-Rainbow gravity.

\subsection{Stellar Structure}

In order to study compact objects, we assume spherical symmetry for the metric. This way, by replacing the usual GR quantities $\widetilde{e}_{i}$ for spherical symmetry into Eq. (\ref{eq3}), we obtain:
\begin{equation}
    ds^{2}=-\frac{B(r)}{\Xi^{2}(x)} dt^{2}+\frac{A(r)}{\Sigma^{2}(x)} dr^{2}+\frac{r^{2}}{\Sigma^{2}(x)}(d\theta^{2}+\sin{\theta}^{2}d\phi^{2}),
    \label{eq5}
\end{equation}
where $A(r)$ and $B(r)$ are radial functions. We can see that our metric depends on the rainbow functions $\Xi(x)$ and $\Sigma(x)$, that is, it depends on the energy of the probe particles. However, note that the coordinates $r$, $t$, $\theta$ and $\phi$ don't have this dependency. 

We assume that stellar matter can be described as an anisotropic fluid represented by the following stress-energy tensor \cite{Arbanil2016,Garattini2016,Sharif2018}:
\begin{equation}
    T_{\mu\nu}=p_{t}g_{\mu\nu}+(p_{t}+\rho)U_{\mu}U_{\nu}+(p_{r}-p_{t})N_{\mu}N_{\nu}
    \label{eq12},
\end{equation}
where $p_{t}(r)$, $\rho(r)$ and $p_{r}(r)$ are respectively the tangential pressure, the energy density and the radial pressure of the fluid. The quantities $U_{\mu}$ and $N_{\mu}$ represent respectively the four velocity and radial unit vector, whose definitions are:
 \begin{equation}
        U^{\mu}=\left( \frac{\Xi(x)}{\sqrt{B(r)}},0,0,0\right)
    \label{eq13},
 \end{equation}
 \begin{equation}
        N^{\mu}=\left( 0,\frac{\Sigma(x)}{\sqrt{A(r)}},0,0\right)
    \label{eq13a}.
 \end{equation}
These quantities obey the conditions: $U_{\nu}U^{\nu}=-1$, $N_{\nu}N^{\nu}=1$ and $U_{\nu}N^{\nu}=0$.

Rewriting Eq. (\ref{eq8}) in its covariant form and rearranging its terms, so that we have the usual Einstein tensor on the left side of the equation, we obtain the usual Einstein equation with an effective energy-momentum tensor, as follows:
 \begin{equation}
     R_{\mu\nu}-\frac{1}{2}g_{\mu\nu}R=8\pi G\tau_{\mu \nu},
     \label{eq13b}
 \end{equation}
 where
 \begin{equation}
     \tau_{\mu\nu}=T_{\mu\nu}-\frac{(1-\lambda)}{2(1-\lambda)}g_{\mu\nu}T.
      \label{eq13c}
 \end{equation}
Now, by using the definitions for the energy dependent spherically symmetric metric given in Eq.(\ref{eq5}) and the energy-momentum tensor for an anisotropic fluid shown in 
Eq.(\ref{eq12}), we calculate the components of the field equations (\ref{eq13b}), and obtain:
\begin{equation}
 -\frac{B}{r^{2}A} + \frac{B}{r^{2}} + \frac{A'B}{rA^{2}} = 8\pi GB \Bar{\rho},  \label{eq14}
\end{equation}

\begin{equation}
-\frac{A}{r^{2}} + \frac{B'}{rB} + \frac{1}{r^2}  = 8\pi GA \Bar{p}_{r},  \label{eq15}
\end{equation}

{\fontsize{9.7}{12}
\begin{align}
 -\frac{B'^{2}r^{2}}{4AB^{2}} - \frac{A'B'r^{2}}{4A^{2}B} + \frac{B'' r^{2}}{2AB} - &\frac{A'r}{2A^{2}} + \frac{B'r}{2AB} \nonumber \\
& = 8\pi Gr^{2} \Bar{p}_{t},  \label{eq16}
\end{align}}
where $\Bar{\rho}$, $\Bar{p}_{r}$ and $\Bar{p}_{t}$ are the effective energy density, the effective radial pressure and the effective tangential pressure respectively, which are defined as
\begin{align}
    \Bar{\rho} & = \frac{1}{\Sigma(x)^{2}}\left[\alpha_{1}\rho+\alpha_{2}p_{r}+2\alpha_{2}p_{t}\right],\label{eq17}\\
    \Bar{p}_{r} & = \frac{1}{\Sigma(x)^{2}}\left[\alpha_{2}\rho+\alpha_{1}p_{r}-2\alpha_{2}p_{t}\right], \label{eq18}\\
    \Bar{p}_{t} & = \frac{1}{\Sigma(x)^{2}}\left[\alpha_{2}\rho-\alpha_{2}p_{r}+\alpha_{3}p_{t}\right], \label{eq18b}
\end{align}
where
\begin{equation*}
    \alpha_{1}=\frac{1-3\lambda}{2(1-2\lambda)}, \qquad \alpha_{2}=\frac{1-\lambda}{2(1-2\lambda)}, \qquad
    \alpha_{3}=-\frac{\lambda}{1-2\lambda}.
\end{equation*}
The field equations obtained here, Eqs. (\ref{eq14})$-$(\ref{eq16}), resemble those achieved in GR for a static spherically symmetric anisotropic fluid. The main difference is that now instead of the usual pressures and energy density, we have the equivalent effective quantities. Thus, similar to what is done in the GR case, we can redefine the function $A(r)$ in terms of a new function $M(r)$ as follows:
\begin{equation}
    A(r)=\left[1-\frac{2GM(r)}{r} \right]^{-1},
    \label{eq19}
\end{equation}
By making this replacement in equation (\ref{eq14}), we can solve the $M(r)$ equation by direct integration so that
\begin{equation}
    M(r)=\int_{0}^{R} 4\pi r'^{2}\Bar{\rho}(r')dr'.
    \label{eq20}
\end{equation}
From the equation above we can conclude that $M(r)$ is the stellar mass, and that $r=R$ is the stellar radius. Note that the mass depends on the effective energy density $\Bar{\rho}$. Thus, as in our previous work, we conclude that the Rastall-Rainbow $\lambda$ and $\Sigma$ parameters modify the mass values. Also, it is important to point out that $\Bar{\rho}$ depends on both the radial pressure $p_{r}$ and the tangential pressure $p_{t}$. Therefore, in the case analyzed here where $p_{r} \neq p_{t}$ the mass values are also affected by the anisotropy.
 We can verify that in the case where $p_{r}=p_{t}$ we recover the definition of $\Bar{\rho}$ from our previous article \cite{mota2019}. Moreover, by assuming $\lambda=1$ and $\Sigma=1$ we find $\Bar{\rho}=\rho$ so that the GR mass definition is retrieved.
 
By calculating the modified conservation law for the energy-momentum tensor $T^{\nu}_{\:\:\:\mu;\nu}=\Bar{\lambda}R_{\mu}$, we get:
\begin{equation}
 \Bar{p}_{r}' = -(\Bar{p}_{r}+\Bar{\rho})\frac{B'}{2B}+2\frac{\Bar{\sigma}}{r},
 \label{eq21}
\end{equation}
where $\Bar{\sigma}=\Bar{p}_{t}-\Bar{p}_{r}$. Rearranging Eq. (\ref{eq15}) and employing the relation shown in Eq.(\ref{eq19}), we find
\begin{align}
 \frac{B'}{2B} & = \frac{M+4\pi G r^{3}\Bar{p}_{r}}{r(r-2M)}. \label{eq22} 
\end{align}
Now, we can replace the equation above into Eq. (\ref{eq21}) to eliminate the function $B$, in order to get:
\begin{equation}
\Bar{p}_{r}' = -(\Bar{p}_{r}+\Bar{\rho})\frac{M+4\pi G r^{3}\Bar{p}_{r}}{r(r-2M)}+2\frac{\Bar{\sigma}}{r}. 
\label{eq23}
\end{equation}
This equation describes the hydrostatic equilibrium equation in the context of Rastall-Rainbow gravity modified by the inclusion of the anisotropy factor $\Bar{\sigma}=\Bar{p}_{t}-\Bar{p}_{r}$. 
To take anisotropy into account, we consider an anisotropic profile dependent radial pressure and geometric quantities that can be derived from space-time geometry \cite{Arbanil2016}, as given by
\begin{equation}
\sigma = \beta p_{r}(1-e^{-\lambda}),
\label{eq24}
\end{equation}
where $\beta$ is the anisotropic constant. The expression $\Bar{\sigma} = \sigma/\Sigma(x)^{2}$ reflects the effects of anisotropy in equation (\ref{eq23}). The function of the metric $e^{-\lambda}$ in equation (\ref{eq24}) is here identified as follows
\begin{equation}
e^{-\lambda} \equiv A(r)^{-1}=\left[1-\frac{2GM(r)}{r} \right].
\end{equation}
In the next section, we use equation (\ref{eq23}) to analyze the gravitational equilibrium of neutron stars.

\section{Results and discussion}

Macroscopic properties, such as mass and radius of anisotropic neutron stars in the context of Rastall-Rainbow gravity are studied next. Initially, it is necessary to construct an appropriate equation of state (EoS) to model the nuclear matter contained within the star. For this purpose, we use a model originated from the relativistic mean field theory (RMF), the IU-FSU parameterization proposed in \cite{iu-fsu}. According to the authors in \cite{mota2019}, in addition to supporting the tests proposed in \cite{Dutra2015}, the IU-FSU also satisfactorily reproduces the recent astrophysical constraint from the observation of GW170817 \cite{Lourenco2018}. We consider two cases for the EoS: first, with nucleonic matter only; second, with the inclusion of the eight lightest baryons, and the results for both cases are shown below in the tables and corresponding figures. 

The usual TOV solutions from general relativity are obtained using $ \Sigma = 1 $, $ \lambda = 1 $ and $ \sigma = 0 $. They are represented by the continuous purple lines in the figures and the resulting values for the maximum mass (and the corresponding radius for this solution) are listed on Tables I to IV. Additionally, we use the full BPS EoS to describe the stellar outer crust \cite{bps}.

We model the anisotropic profile $\sigma$ according to equation (\ref{eq24}), which has been previously studied in the context of GR  \cite{Arbanil2016}.  Note that effects due to the anisotropy vanish on the stellar surface, namely $\sigma (R) = 0$. For the  parameters of the Rastall-Rainbow gravity, we use the same values considered in \cite{mota2019}. As for the anisotropy parameter $ \beta $, we have tested a range of positive and negative values: $ \beta = 0, \pm 0.5, \pm 1.0, \pm 1.5, \pm 2.0, \pm 2.5. $. We observe that negative values drastically decrease the maximum stellar mass, and, while we maintain one such example in each table for pedagogical reasons, they are effectively discarded. In contrast, when combined with the effects of Rastall-Rainbow, we find that the positive values for $ \beta $ provide the best results for the mass-radius of a neutron star family when confronted with modern astrophysical constraints: the neutron star in the quiescent low-mass X-ray binary (LMXB) NGC 6397 \cite{Grindlay2001,Guillot2010,Heinke2014} and the extremely massive millisecond pulsar MSP J0740+6620 \cite{Cromartie2019}. Therefore, the positive values have been further explored.

\begin{figure*}[t]
\centering
\begin{tabular}{ll}
\includegraphics[width=5.7cm,angle=270]{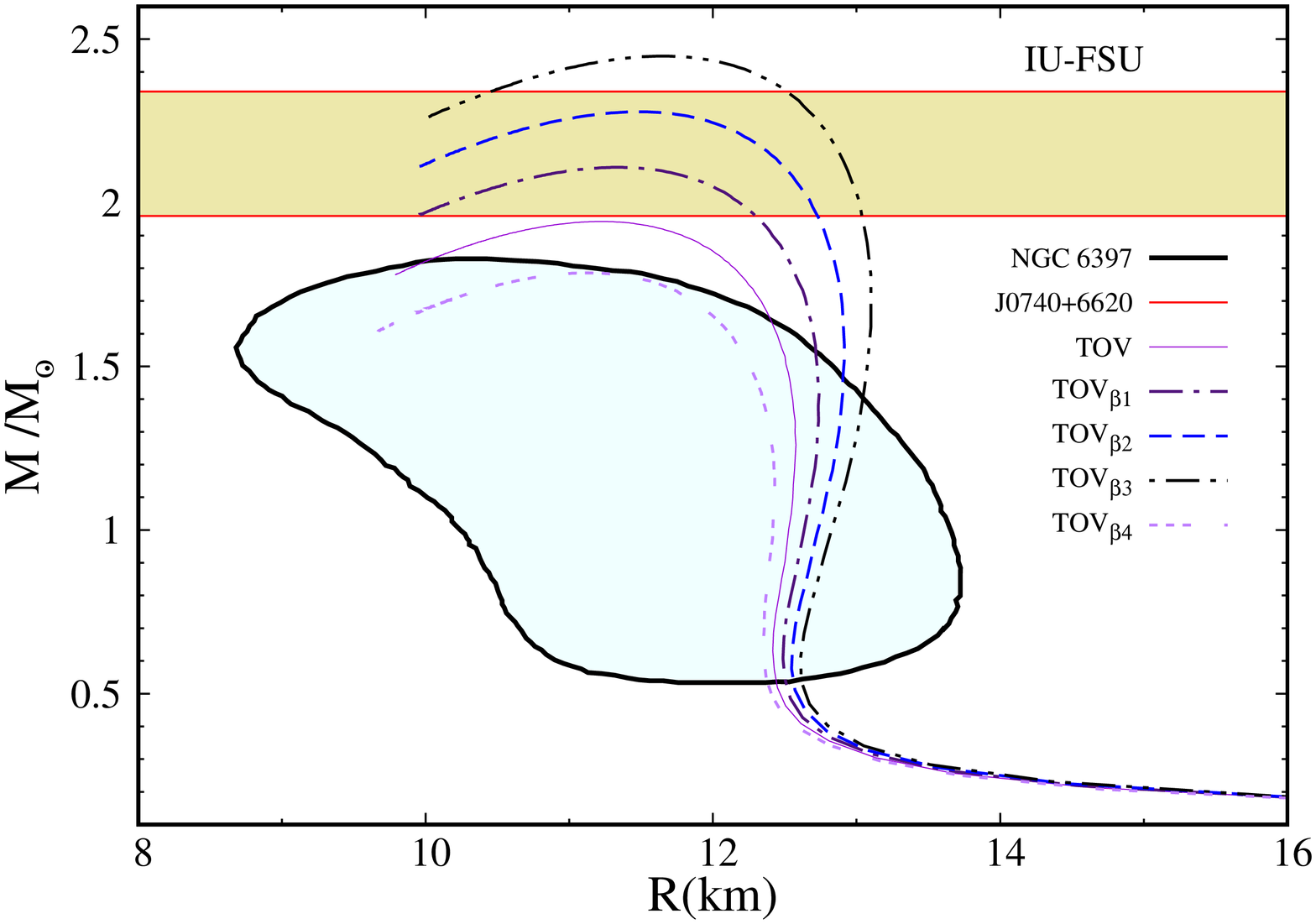}
\end{tabular}
\caption{Mass-radius relation for a family of hadronic stars described with the IU-FSU EoS without hyperons. We analyze the effects caused by varying the anisotropic parameter $\beta$. 
}
\label{fig1}
\end{figure*}

First, we analyze the effects of the proposed anisotropic profile on solutions of the usual TOV equations from GR. The IU-FSU EoS without hyperons yields a maximum mass slightly smaller than $ 2.0M_{\odot} $ and the radius corresponding to the canonical star ($ 1.4 M_{\odot} $) satisfies the restrictions imposed by GW170817, which suggests, for this star, a radius value between 10.5 - 13.4 km \cite{PhysRevLett.119.161101}. From Table \ref{tab_TOV} and Fig. \ref{fig1} one can see that, while the anisotropic effect is to increase the maximum mass, the corresponding radius also increases as the mass grows, i.e., as $\beta$ assumes larger values, maximum masses and radii increase. Our tests indicate that the results are extremely sensitive to these variations and therefore we have restricted a limiting value to this parameter around $\beta = 2.0$.

\begin{figure*}[t]
\centering
\begin{tabular}{ll}
\includegraphics[width=5.7cm,angle=270]{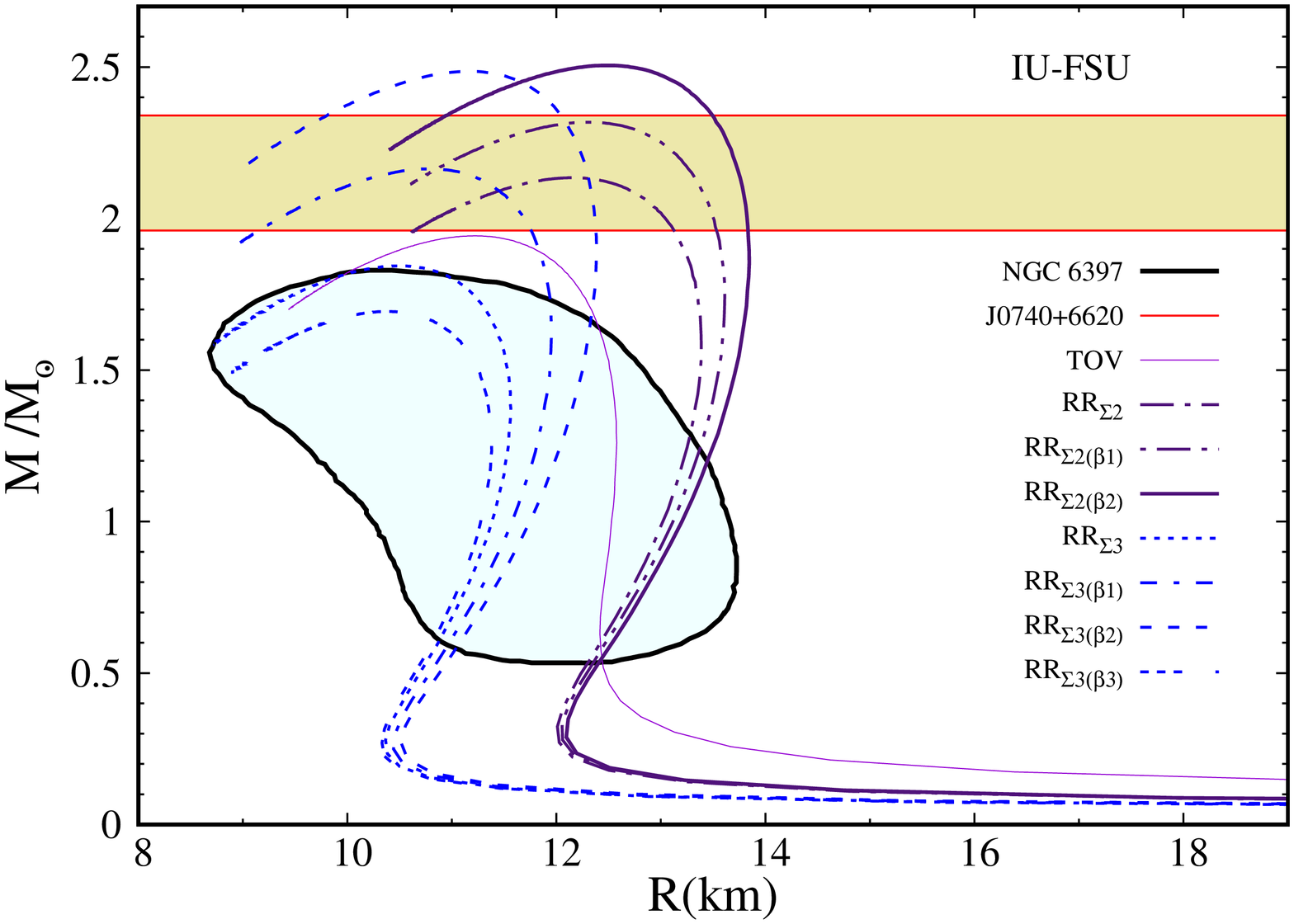}
\includegraphics[width=5.7cm,angle=270]{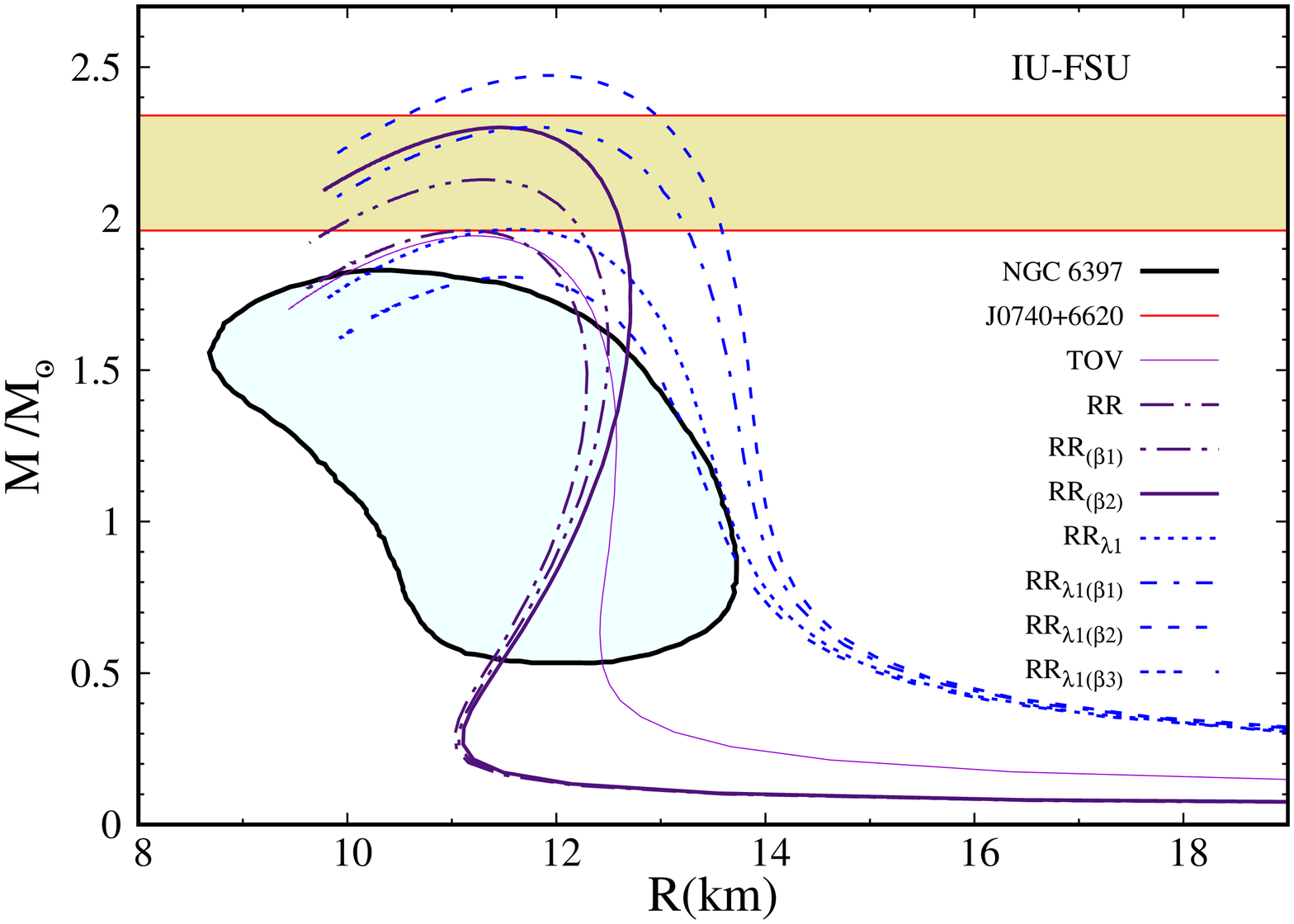} 
\end{tabular}
\caption{Mass-radius relation for a family of hadronic stars described with the IU-FSU EoS without hyperons. We analyze the effects caused by varying the Rainbow and anisotropic parameters `` $\Sigma$, $\beta$ '' (left) while keeping the other parameter fixed
and, the effects of varying the Rastall and anisotropic parameters `` $\lambda$, $\beta$ '' (right) while keeping $\Sigma$ fixed. 
}
\label{fig2}
\end{figure*}

Next, by using some fixed value settings for the Rastall-Rainbow theory, we again analyze the effect of the anisotropy on the neutron star mass-radius relation and the results are shown in Table \ref{tab_IUFSU}. We vary the anisotropy parameter around the values mentioned above. Although the Rastall theory hardly affects the maximum stellar mass and dramatically increases the corresponding radius, the Rainbow theory, on the other hand, increases both depending on the values chosen \cite{mota2019}. Notice in  Fig. \ref{fig2} that the implementation of the anisotropy makes it possible to obtain larger maximum mass configurations as compared to the results obtained in \cite{mota2019}. The combination of both theories (Rastall and rainbow) plus the introduction of the anisotropy allows the maximum mass to increase substantially to satisfy modern astrophysical observations.

Due to the extreme density conditions within neutron stars, the possibility of the appearance of baryonic species more massive than protons and neutrons, the so-called hyperons, cannot be discarded. However, their inclusion in the nuclear model is known to soften the EoS and therefore to produce smaller  maximum masses than the ones produced by their nucleonic counterpart. This feature is known as the hyperon puzzle, a hot topic in nuclear astrophysics. There are different ways to 
circumvent this problem. One of them is the inclusion of new degrees of freedom with strangeness content \cite{Lopes2018}, another is the solution of TOV-like equations in the braneworld \cite{benito2014} and still one more possibility is the introduction of the caotic field approximation \cite{Lopes2019}. In the present work we show that the problem can also be solved by including anisotropic effects on neutron stars in the context of Rastall-Rainbow gravity.

\begin{figure*}[t]
\centering
\begin{tabular}{ll}
\includegraphics[width=5.7cm,angle=270]{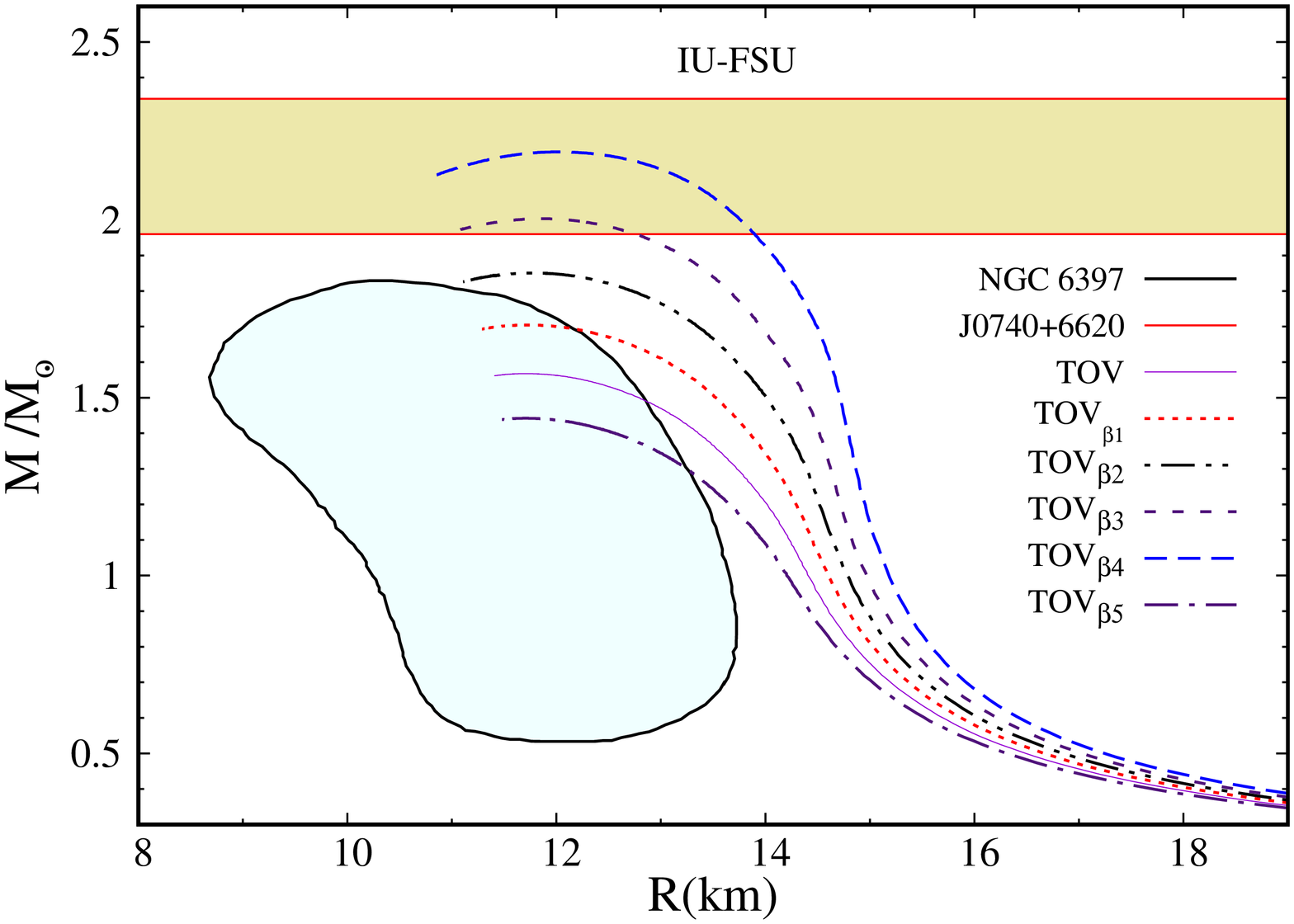}
\end{tabular}
\caption{Mass-radius relation for a family of hadronic stars described with the IU-FSU EoS with hyperons. We analyze the effects caused by varying the anisotropic parameter $\beta$. 
}
\label{fig4}
\end{figure*}

\begin{figure*}[t]
\centering
\begin{tabular}{ll}
\includegraphics[width=5.7cm,angle=270]{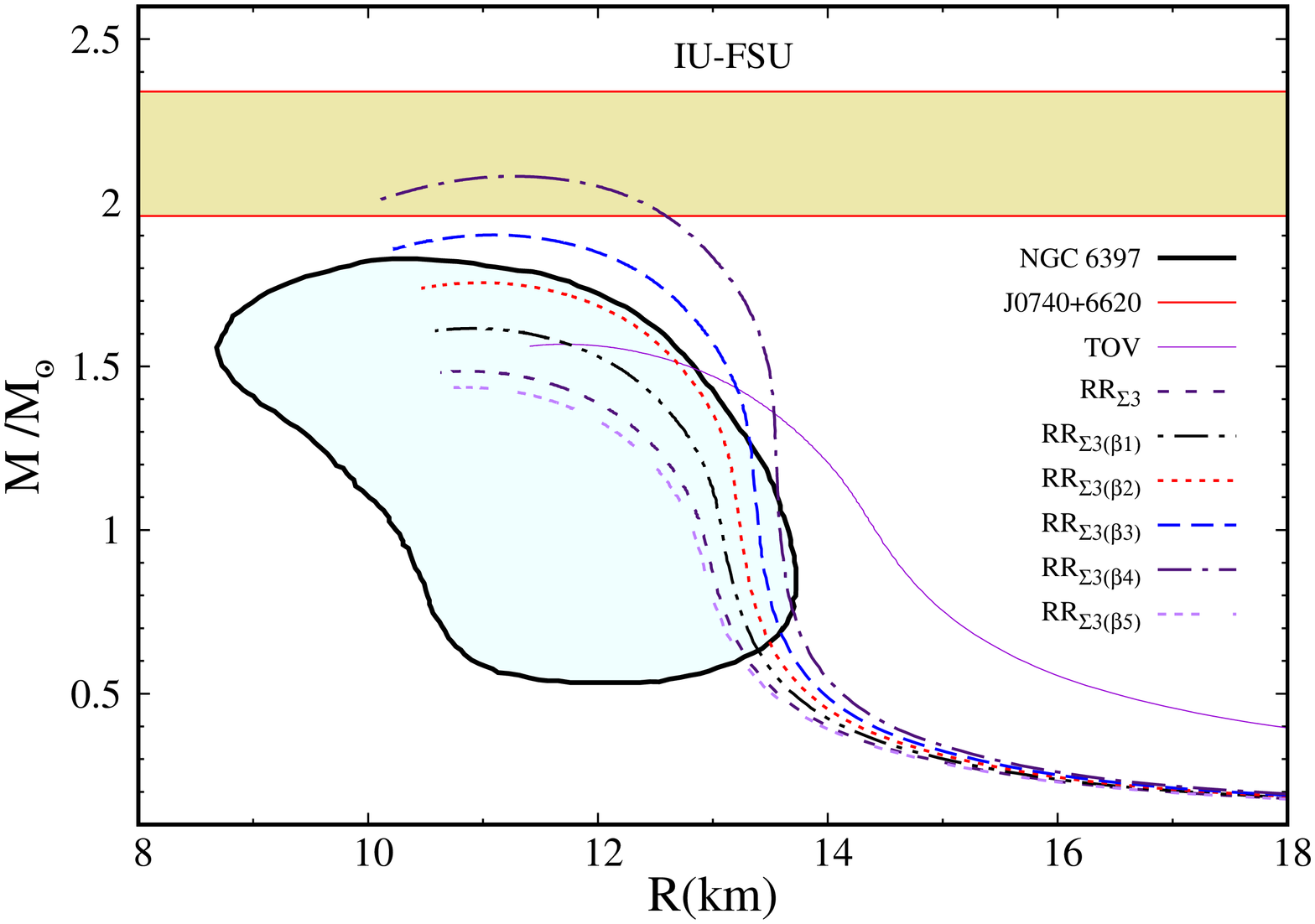}
\includegraphics[width=5.7cm,angle=270]{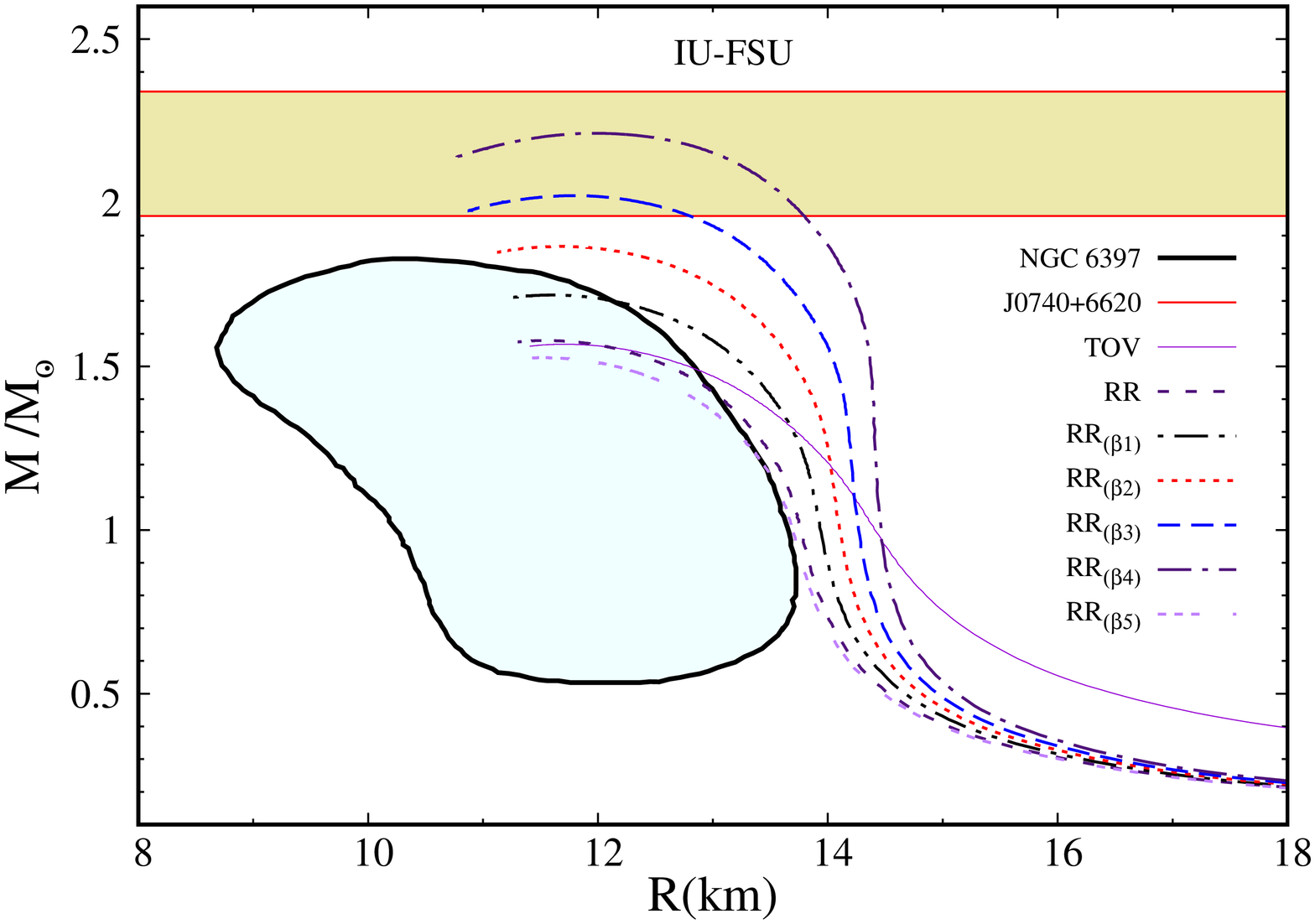} 
\end{tabular}
\caption{Mass-radius relation for a family of hadronic stars described with the IU-FSU EoS with hyperons. We analyze the effects caused by varying the Rainbow and anisotropic parameters `` $\Sigma$, $\beta$ '' (left) while keeping the other parameter fixed
and, the effects of varying the Rastall and anisotropic parameters `` $\lambda$, $\beta$ '' (right) while keeping $\Sigma$ fixed. 
}
\label{fig5}
\end{figure*}

Once again, we start by analyzing the effects of the anisotropy on the usual TOV equations from GR. The results are displayed in Table \ref{tab_IUFSU_TOV} and Fig. \ref{fig4}. Again, as the maximum mass grows, the radius for the canonical star also grows and therefore NGC 6397 constraint is not satisfied. In Table \ref{tab_IUFSU_RR_anis} and Fig. \ref{fig5}, we show the effects of Rastall-rainbow gravity on the properties of anisotropic neutron stars obtained with IU-FSU EoS with hyperons. As for the meson-hyperon coupling constants, we use the same values as in \cite{Glendenning1991}. We are now able to adjust values for both NGC 6397 and MSP J0740+6620 observations.

\section{Concluding Remarks}

In this work we have studied the equilibrium configuration of anisotropic neutron stars using one typical equation of state (IU-FSU) for standard hadronic matter, which  was selected  in order to agree with bulk nuclear matter properties, the neutron star maximum masses recently observed and constraints from gravitational wave observations. After we have chosen the equation of state we proceeded to calculate the stellar structure and, for this purpose, we used the well know TOV equations where general relativity is considered and afterwards we employed the modified version of the TOV equations for the case of Rastall-Rainbow gravity.

After this initial tests we have come up with an exploration of the parameter window of RR gravity. To avoid very arbitrary parameter values, we have chosen the same values obtained in the previous 
paper with which both massive and small radii stars could be described \cite{mota2019}. We have then explored anisotropic effects both on stellar matter constituted of nucleons only and also on hyperonic matter as input to RR equations. We observed that neutron star masses can extend to values next to 2.5 M$_{\odot}$. As already pointed out in \cite{mota2019}, while the maximum mass increases, the RR theory allows the decrease of the canonical NS radii.

We have shown two main global effects. At first, when only anisotropy is considered in the usual GR framework, it can significantly alter the maximum stellar mass. On the other hand, when Rastall-Rainbow gravity and anisotropy are considered simultaneously, the positive values for $\beta$ provide the best results for the radius and mass of a neutron star, which satisfy recent astrophysical constraints. Moreover, although the inclusion of hyperons in the nuclear model results in a softer EoS and smaller stellar masses than those produced by their nucleic counterparts, we have shown that the problem can be solved by including anisotropic effects on neutron stars in the context of Rastall-Rainbow gravity. Within this new perspective, we can produce macroscopic properties within the currently accepted range for a variety of parameters when combining the effects of RR with anisotropic stellar configurations.

Finally, it is worth mentioning that all our results were motivated by previous works \cite{Hendi2016,mota2019} in which the
dependence of the Rainbow functions with the energy scale of the star
was not considered, but rather taken as effective parameters. This consideration is justified because the energy dependence of the previously mentioned functions is not very well know. This technical detail can be traduced in a more complex form of the resulting field equations and we plan to work in this problem in a future work.

\acknowledgments
\noindent This work is a part of the project CNPq-INCT-FNA Proc. No. 464898/2014-5. D.P.M. acknowledges partial support from CNPq (Brazil) under grant 301155/2017-8, C.E.M., F.M.S. and C.V.F. are supported respectively by Capes (Brazil) scholarship and PNPD program and L.C.N.S. would like to thank Conselho Nacional de Desenvolvimento Cient\'ifico e Tecnol\'ogico (CNPq) for partial financial support through the research Project No. 155361/2018-0.


\begin{table*}[t]
\centering
\caption{Macroscopic properties for different values of the $\beta$  parameter corresponding to the mass-radius diagram in FIG.\ref{fig1}.}
{\small
\begin{tabular}{c|c|c|ccccccc}
\hline
\hline
\textbf{General Relativity} & Model & \ \ TOV \ \ & TOV$_{\beta1}$ \ \ & TOV$_{\beta2}$ \ \ &TOV$_{\beta3}$ \ \ & TOV$_{\beta4}$ \tabularnewline 

\hline
 Parameter
 
& $\boldsymbol{\beta}$  &  \textbf{0.0}  &  \textbf{0.5} &  \textbf{1.0} &  \textbf{1.5}  & \textbf{-0.5}  \tabularnewline

\hline
& $M_{max}$  &  1.94 $M_\odot$ &  2.10 $M_\odot$  & 2.27 $M_\odot$  &  2.44 $M_\odot$  & 1.78 $M_\odot$ \tabularnewline

IU-FSU & $R_{M_{max}}$ &  11.22 km  \ \ &  11.33 km   \ \ & 11.48 km \ \ & 11.65 km  \ \ & 11.12 km  \tabularnewline

 &$R_{1.4}$& 12.55 km  \ \ &  12.74 km \ \ & 12.90 km  \ \ & 13.04 km  \ \ & 12.34 km  \tabularnewline
\hline
\hline
\label{tab_TOV}
\end{tabular}
}
\end{table*}

\begin{table*}[t]
\centering
\caption{Macroscopic properties for different values of the $\lambda$, $\Sigma$ and $\beta$ parameters corresponding to the mass-radius diagram in FIG.\ref{fig2}.}
{\small
\begin{tabular}{c|c|c|ccccccc}
\hline
\hline
\textbf{Rainbow} & Model & \ \ TOV \ \ & RR$_{\Sigma2}$ \ \cite{mota2019} \ \ & RR$_{\Sigma2(\beta1)}$ \ \ & RR$_{\Sigma2(\beta2)}$ \ \ & RR$_{\Sigma3}$ \ \cite{mota2019} \ \ & RR$_{\Sigma3(\beta1)}$ \ \ & RR$_{\Sigma3(\beta2)}$ \ \ & RR$_{\Sigma3(\beta3)}$ \tabularnewline 

\hline
& $\mathbf{\Sigma}$  &  \textbf{1.0}  & \textbf{1.1}  & \textbf{1.1} & \textbf{1.1}  &  \textbf{0.95}  & \textbf{0.95} & \textbf{0.95}  & \textbf{0.95}  \tabularnewline
 
Parameters & $\boldsymbol{\beta}$  &  \textbf{0.0}  &  \textbf{0.0} &  \textbf{0.5} &  \textbf{1.0}  & \textbf{0.0}  &  \textbf{1.0} & \textbf{2.0} & \textbf{-0.5}  \tabularnewline
 
&$\lambda$  &  1.0  &  0.999 &  0.999 &  0.999  & 0.999  &  0.999 & 0.999 & 0.999  \tabularnewline

\hline
& $M_{max}$  &  1.94 $M_\odot$ &  2.13 $M_\odot$  & 2.31 $M_\odot$  &  2.50 $M_\odot$  & 1.84 $M_\odot$ & 2.16 $M_\odot$ & 2.48 $M_\odot$ & 1.69 $M_\odot$\tabularnewline

IU-FSU & $R_{M_{max}}$ &  11.22 km  \ \ &  12.16 km   \ \ & 12.30 km \ \ & 12.48 km  \ \ & 10.49 km  \ \  & 10.78 km  \ \ & 11.17 km \ \ & 10.38 km \ \ \tabularnewline

 &$R_{1.4}$& 12.55 km  \ \ &  13.34 km \ \ & 13.49 km  \ \ & 13.64 km  \ \ & 11.55 km  \ \ & 11.89 km  \ \ & 12.17 km \ \ & 11.34 km \ \ \tabularnewline
\hline
 &  &   \ \ &    \ \ &\ \ &   \ \ &    \ \  &   \ \ &  \ \ \tabularnewline

 &  &  General  \ \ &    \ \ &\ \ &   \ \ & Modified   \ \  &   \ \ &  \ \ \tabularnewline
  &  &  Relativity  \ \ &    \ \ &\ \ &   \ \ & Gravity \ \  &   \ \ &  \ \ \tabularnewline
   &  &    \ \ &     \ \ &\ \ &   \ \ &    \ \  &   \ \ &  \ \ \tabularnewline

\hline
\textbf{Rastall} & Model & \ \ TOV \ \ \ & RR \ \cite{mota2019} \ \ & RR$_{(\beta1)}$ \ \ \ & RR$_{(\beta2)}$ \ \ \ & RR$_{\lambda1}$ \ \cite{mota2019} \ \ \ & RR$_{\lambda1(\beta1)}$ \ \ \ & RR$_{\lambda1(\beta2)}$ \ \ \ & RR$_{\lambda1(\beta3)}$ \tabularnewline 
\hline
& $\boldsymbol{\lambda}$  &  \textbf{1.0}  &  \textbf{0.999} &  \textbf{0.999}  &  \textbf{0.999}  & \textbf{1.001} & \textbf{1.001} & \textbf{1.001}  & \textbf{1.001} \tabularnewline

Parameters & $\boldsymbol{\beta}$  &  \textbf{0.0}  &  \textbf{0.0} &  \textbf{0.5} &  \textbf{1.0}  & \textbf{0.0}  &  \textbf{1.0} & \textbf{1.5} & \textbf{-0.5}  \tabularnewline

& $\Sigma$  &  1.0  & 1.01 & 1.01 & 1.01  & 1.01 &  1.01  & 1.01 & 1.01  \tabularnewline

\hline
& $M_{max}$  &  1.94 $M_\odot$ &  1.96 $M_\odot$  & 2.12 $M_\odot$  &  2.30 $M_\odot$  &  1.96 $M_\odot$ & 2.30 $M_\odot$  & 2.47 $M_\odot$ & 1.80 $M_\odot$\tabularnewline

IU-FSU & $R_{M_{max}}$ &  11.22 km  \ \ &  11.16 km   \ \ & 11.29 km \ \ & 11.45 km  \ \ & 11.59 km  \ \  & 11.80 km  \ \ & 11.94 km \ \ & 11.55 km \ \ \tabularnewline

 &$R_{1.4}$& 12.55 km  \ \ &  12.28 km \ \ & 12.45 km  \ \ & 12.60 km  \ \ & 13.32 km  \ \ & 13.70 km  \ \ & 13.86 km \ \ & 13.09 km \ \ \tabularnewline
\hline
\hline
\label{tab_IUFSU}
\end{tabular}
}
\end{table*}


\begin{table*}[t]
\centering
\caption{Macroscopic properties for different values of the $\beta$ parameter corresponding to the mass-radius diagram in FIG.\ref{fig4}.}
{\small
\begin{tabular}{c|c|c|ccccccc}
\hline
\hline
\textbf{General Relativity} & Model & \ \ TOV \ \ & TOV$_{\beta1}$ \ \ & TOV$_{\beta2}$ \ \ &TOV$_{\beta3}$  \ \ &TOV$_{\beta4}$ \ \ & TOV$_{\beta5}$ \tabularnewline 

\hline
 Parameter
 
& $\boldsymbol{\beta}$  &  \textbf{0.0}  &  \textbf{0.5} &  \textbf{1.0} &  \textbf{1.5} &  \textbf{2.1} & \textbf{-0.5}  \tabularnewline

\hline
& $M_{max}$  &  1.56 $M_\odot$ &  1.70 $M_\odot$  & 1.85 $M_\odot$  &  2.00 $M_\odot$  & 2.19 $M_\odot$ & 1.44 $M_\odot$ \tabularnewline

IU-FSU & $R_{M_{max}}$ &  11.70 km  \ \ &  11.72 km   \ \ & 11.78 km \ \ & 11.85 km  \ \ & 12.01 km  \ \ & 11.69 km \tabularnewline

 &$R_{1.4}$& 13.34 km  \ \ &  13.85 km \ \ & 14.22 km  \ \ & 14.51 km  \ \ & 14.72 km  \ \ & 12.56 km  \tabularnewline
\hline
\hline
\label{tab_IUFSU_TOV}
\end{tabular}
}
\end{table*}


\begin{table*}[t]
\centering
\caption{Macroscopic properties for different values of the $\lambda$, $\Sigma$ and $\beta$ parameters corresponding to the mass-radius diagram in FIG.\ref{fig5}.}
{\small
\begin{tabular}{c|c|c|ccccccc}
\hline
\hline
\textbf{Rainbow} & Model & \ \ TOV \ \ & RR$_{\Sigma3}$ \ \ & RR$_{\Sigma3(\beta1)}$ \ \ & RR$_{\Sigma3(\beta2)}$ \ \ & RR$_{\Sigma3(\beta3)}$ \ \ & RR$_{\Sigma3(\beta4)}$ \ \ & RR$_{\Sigma3(\beta5)}$ \tabularnewline 

\hline
& $\boldsymbol{\beta}$  &  \textbf{0.0}  &  \textbf{0.0} &  \textbf{0.5} &  \textbf{1.0}  & \textbf{1.5}  &  \textbf{2.1} & \textbf{-0.2} \tabularnewline

Parameters & $\Sigma$  &  1.0  & 0.95  & 0.95 & 0.95  &  0.95  & 0.95 & 0.95 \tabularnewline 
 
&$\lambda$  &  1.0  &  0.999 &  0.999 &  0.999  & 0.999  &  0.999 & 0.999 \tabularnewline

\hline
& $M_{max}$  &  1.56 $M_\odot$ &  1.48 $M_\odot$  & 1.61 $M_\odot$  &  1.75 $M_\odot$  & 1.90 $M_\odot$ & 2.08 $M_\odot$ & 1.43 $M_\odot$ \tabularnewline

IU-FSU & $R_{M_{max}}$ &  11.70 km  \ \ &  10.85 km   \ \ & 10.91 km \ \ & 10.98 km  \ \ & 11.08 km  \ \  & 11.25 km  \ \ & 10.84 km \ \ \tabularnewline

 &$R_{1.4}$& 13.34 km  \ \ &  11.92 km \ \ & 12.51 km  \ \ & 12.92 km  \ \ & 13.23 km  \ \ & 13.52 km  \ \ & 11.54 km \ \ \tabularnewline
\hline
 &  &   \ \ &    \ \ &\ \ &   \ \ &    \ \  &   \ \ &  \ \ \tabularnewline

 &  &  General  \ \ &    \ \ &\ \ &   \ \ & Modified   \ \  &   \ \ &  \ \ \tabularnewline
  &  &  Relativity  \ \ &    \ \ &\ \ &   \ \ & Gravity \ \  &   \ \ &  \ \ \tabularnewline
   &  &    \ \ &     \ \ &\ \ &   \ \ &    \ \  &   \ \ &  \ \ \tabularnewline

\hline
\textbf{Rastall} & Model & \ \ TOV \ \ \ & RR \ \ & RR$_{(\beta1)}$ \ \ \ & RR$_{(\beta2)}$ \ \ \ & RR$_{(\beta3)}$ \ \ \ & RR$_{(\beta4)}$ \ \ \ & RR$_{(\beta5)}$ \tabularnewline 
\hline
& $\boldsymbol{\beta}$  &  \textbf{0.0}  &  \textbf{0.0} &  \textbf{0.5} &  \textbf{1.0}  & \textbf{1.5}  &  \textbf{2.1} & \textbf{-0.2} \tabularnewline

Parameters & $\lambda$  &  1.0  &  0.999 &  0.999  &  0.999  & 0.999 & 0.999 & 0.999 \tabularnewline

& $\Sigma$  &  1.0  & 1.01 & 1.01 & 1.01  & 1.01 &  1.01  & 1.01 \tabularnewline

\hline
& $M_{max}$  &  1.56 $M_\odot$ &  1.57 $M_\odot$  & 1.71 $M_\odot$  &  1.86 $M_\odot$  &  2.02 $M_\odot$ & 2.21 $M_\odot$  & 1.52 $M_\odot$ \tabularnewline

IU-FSU & $R_{M_{max}}$ &  11.70 km  \ \ &  11.53 km   \ \ & 11.59 km \ \ & 11.68 km  \ \ & 11.79 km  \ \  & 11.97 km  \ \ & 11.52 km \ \  \tabularnewline

 &$R_{1.4}$& 13.34 km  \ \ &  13.07 km \ \ & 13.51 km  \ \ & 13.87 km  \ \ & 14.14 km  \ \ & 14.39 km  \ \ & 12.83 km \ \ \tabularnewline
\hline
\hline
\label{tab_IUFSU_RR_anis}
\end{tabular}
}
\end{table*}

\bibliography{referencias}
\end{document}